\documentclass[twocolumn,prl,groupedaddress,showpacs,superscriptaddress,amsmath]{revtex4}
\usepackage{color}
\usepackage{epsfig,amsfonts,amsbsy,graphicx,subfigure}

\begin{document}

\title{Entropy production of nano systems with timescale separation}
 
 \author{Shou-Wen Wang}
\affiliation{Beijing Computational Science Research Center, Beijing, 100094, China}
\affiliation{Department of Engineering Physics, Tsinghua University, Beijing, 100086, China}
\author{Kyogo Kawaguchi}
\affiliation{Department of Systems Biology, Harvard Medical School, Boston, MA 02115, USA}
\author{Shin-ichi Sasa}
\affiliation{Department of Physics, Kyoto University, Kyoto 606-8502, Japan}
\author{Lei-Han Tang}
\affiliation{Beijing Computational Science Research Center, Beijing, 100094, China}
\affiliation{Department of Physics and Institute of Computational and Theoretical Studies, Hong Kong Baptist University, Hong Kong, China}

\date{\today}
  \graphicspath{{./figure/}}
  
\begin{abstract}

Energy flows in bio-molecular motors and machines are vital to their function.  
Yet experimental observations are often limited to a small subset of variables that participate in energy
transport and dissipation. Here we show, through a solvable Langevin model, that the seemingly hidden entropy
production is measurable through the violation spectrum of the
fluctuation-response relation of a slow observable. For general Markov
systems with timescale separation, we prove that the violation
spectrum exhibits a characteristic plateau in the intermediate
frequency region. Despite its vanishing height, the plateau can account for energy dissipation over
a broad timescale.  Our findings suggest a general possibility to probe
hidden entropy production in nano systems without direct observation of fast variables.

\end{abstract}

\maketitle

\emph{Introduction.---}  
Recent advances in technology have made it possible to investigate energetic aspects of an open nano system 
experimentally. Studies have been carried out to quantify, e.g., the energy conversion efficiency of molecular motors~\cite{noji1997direct,toyabe2010nonequilibrium}, 
the information-energy conversion efficiency of an artificial Maxwell demon~\cite{sagawa2010generalized, toyabe2010experimental, berut2012experimental,Koski2015MaxwellDemon}, the entropy production of a quantum tunneling device in a temperature  gradient~\cite{koski2013distribution}, and the effective temperature of single molecules in nonequilibrium steady-states~\cite{dieterich2015single}.   
 
In the most interesting cases, such small systems  are capable of complex dynamic behavior by virtue of multiple degrees of freedom over a broad range of timescales, and furthermore by operating out of equilibrium. 
 The functional features of these systems are usually associated with slow processes, but
recent theoretical studies have provided several examples of ``hidden entropy production'' arising from  
non-equilibrium coupling between fast and slow variables~\cite{hondou2000unattainability,esposito2012stochastic,celani2012anomalous,kawaguchi2013fluctuation,nakayama2015invariance,Chun2015fast,Esposito2015Stochastic, shouwen2015adaptation,Pablo2015Adaptation,bo2014entropy,puglisi2010entropy}.  
A better understanding of the conditions and key characteristics of this phenomenon is desirable
not only from a theoretical viewpoint, but also for developing experimental protocols to uncover 
 channels  of energy dissipation without a full characterization of the possibly great many fast processes
in a complex nano system.

In this Letter, we show that it is indeed possible to trace the hidden entropy production by quantifying the violation 
spectrum of the fluctuation-response relation (FRR) of a slow observable.  
In equilibrium systems, the FRR states that the spontaneous fluctuation of an observable decays 
in the same way as the deviation produced by an external perturbation.   
For a velocity observable $\dot{x}$ which is of particular interest here, the former can be measured via the autocorrelation 
function $C_{\dot{x}}(t)\equiv \langle [\dot{x}(t)- \langle\dot{x}\rangle_{s} ] [\dot{x}(0)-\langle\dot{x}\rangle_{s}] \rangle_{s}$ 
in the stationary ensemble $\langle \cdot\rangle_{s}$,
while the latter is captured by the dynamic response $R_{\dot{x}}(t)\equiv \delta \langle \dot{x}(t)\rangle/\delta h$, 
with $h$ being the perturbing field~\cite{kubo1966fluctuation}.
In frequency, $\tilde{C}_{\dot{x}}(\omega)=2T\tilde{R}_{\dot{x}}'(\omega)$,  
where $T$ is the temperature of the bath and the prime on $\tilde{R}_{\dot{x}}$ denotes its real part.
However, the two properties are not simply related for nonequilibirum systems  
\cite{cugliandolo1997fluctuation,speck2006restoring,baiesi2009fluctuations,seifert2010fluctuation}. 
One significant result achieved in this direction is an equality derived by Harada and Sasa (HS) in Langevin systems.   
The equality connects the steady-state dissipation rate through 
 the frictional motion of $x$,  denoted as $J_x$,  to the integral of the frequency-resolved FRR violation~\cite{harada2005equality,harada2006energy}, 
\begin{equation}
J_x=\gamma  \left \{ \langle \dot{x} \rangle_{s}^2+ \int_{-\infty}^\infty \frac{d\omega}{2\pi} [\tilde{C}_{\dot{x}}(\omega)-2T\tilde{R}_{\dot{x}}'(\omega)] \right \},
\label{eq:HS}
\end{equation}
where $\gamma$ is the friction coefficient.
This equality has been validated experimentally in a driven colloidal system~\cite{toyabe2007experimental}, 
and applied to F$_1$-ATPase, a biomolecular motor, to infer the dissipation rate of the rotary motion~\cite{toyabe2010nonequilibrium,kawaguchi2014nonequilibrium}.

The known applications of the HS equality are for systems that dissipate energy on a slow timescale. 
Here we show that the HS equality can also be used to probe hidden
entropy production that takes place on timescales faster than the relaxation time of the observed variable.
We first demonstrate, in a solvable Langevin model, the use of Eq.~(\ref{eq:HS}) to fully account for the
dissipated energy in a nonequilibrium steady-state.
Interestingly, the FRR violation becomes vanishingly small for large timescale separation, 
while its integral with respect to frequency remains finite.
 We present a proof that this feature of the FRR violation spectrum, arising from
the dissipative coupling between slow and fast variables, is generic for a timescale separated Markov system. 
Based on these findings, we suggest an experimental method to detect  
hidden entropy production from the fluctuation-response spectrum of a slow variable. 

\begin{figure}
\centering
\includegraphics[width=8.5cm]{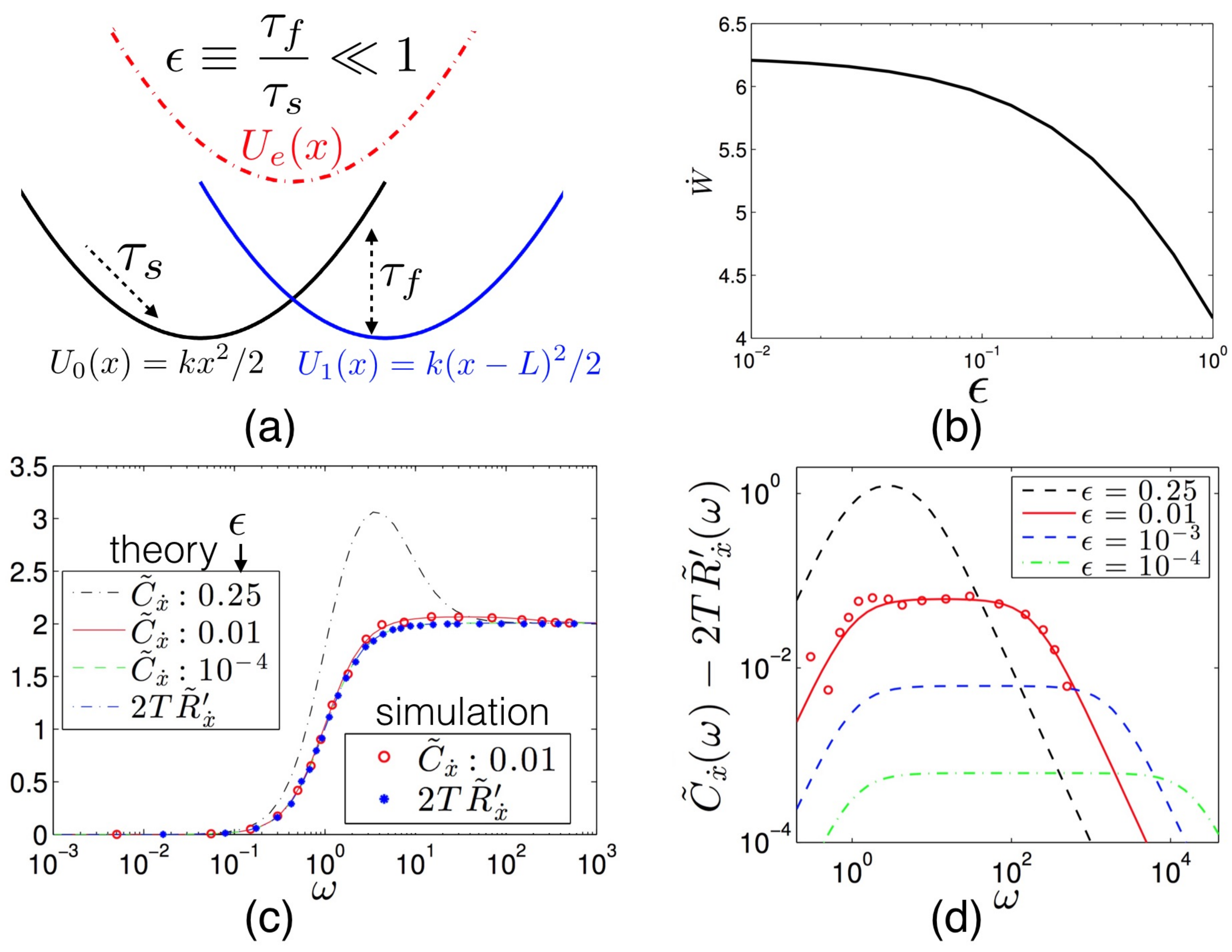}
\caption{(a)  The potential switching model with a fast switching time constant $\tau_f=1/r$ and 
a slow relaxation time constant $\tau_s=\gamma/k$. 
The effective potential $U_e(x)$ in the fast switching limit $\epsilon\ll 1$ is shown.  (b)  Rate of
energy input against $\epsilon$.  (c)  Frequency spectra of the velocity correlation and response functions at 
various values of $\epsilon$.   (d) Frequency spectra of the FRR violation whose
integral yields the hidden entropy production that balances energy input in (b).  
Parameters: $k=\gamma=1$ and $L=5$.  
 Open circles and stars in (c) and (d) give results from a simulated trajectory of length $T_{sp}=10^4\tau_s$ and sampling rate 
 $2\tau_f^{-1}$. The response function is reconstructed from data using a perturbation strength  $h=0.5$. (See Supplemental Material~\cite{supp-hiddenEntropy}.)  }
\label{fig:violation-spectrum}
\end{figure}

\emph{Potential switching model.---}  Consider the one-dimensional, over-damped motion of a bead subjected to a 
potential that switches stochastically between $U_0(x)$ and $U_1(x)$ at a rate $r$.  
Figure~\ref{fig:violation-spectrum}(a) illustrates an experimental realization using a laser trap that
produces a harmonic potential $U_0(x)=kx^2/2$ whose center position switches back and forth between $x= 0$ and $L$ 
[therefore $U_1(x)=k(x-L)^2/2$]~\cite{toyabe2007experimental,dieterich2015single}.   
This can also be viewed  as a minimum Langevin model to study molecular machines which have  fast  binding/unbinding of currency molecules  that triggers transition between different chemical states.
With the  potential state $\sigma_t$ $(=0,1)$ at time $t$,  
and under a perturbing force $h$,  the Langevin equation for the bead position $x$ takes the form,
\begin{equation}
\gamma \dot{x}=-\partial_x U_{\sigma_t}(x)+h(t)+\eta(t). \label{eq:noise}
\end{equation}
Here $\gamma$ is the friction coefficient and $\eta (t)$ is the thermal noise satisfying $\langle \eta(t)\rangle=0$ and 
$\langle \eta(t)\eta(t')\rangle=2\gamma T\delta (t-t')$,  with $T$ the temperature of the bath. 
The Boltzmann constant $k_B$ is set to 1.  

The model defined above has two timescales, $\tau_s=\gamma/k$ for
relaxation within a given potential and $\tau_f=1/r$ for potential switching.
We introduce  $\epsilon \equiv \tau_f/\tau_s$  to characterize the timescale separation between the two competing processes. 
In the fast switching limit $\epsilon\to 0$,  it is straightforward to show that the steady-state distribution
of the bead position takes the Boltzmann form $P^{s}(x)\sim \exp[-U_e(x)/T]$, where $U_e=[U_0(x)+U_1(x)]/2$ 
is the effective potential seen by the bead.  Nevertheless,
due to potential switching, energy is continuously injected into the system at an average rate 
$$\dot{W}=\int_{-\infty}^\infty  r[P_0^{s}(x)-P_1^{s}(x)][U_1(x)-U_0(x)]dx,$$ 
where $P_\sigma^{s}(x)$  is the stationary distribution in the full state space $(x,\sigma)$. 
By analyzing the Fokker-Planck equation in the steady-state satisfied by $P_\sigma^{s}(x)$,
we obtain
\begin{equation}
 \dot{W} \xrightarrow{\epsilon \to 0}  \frac{1}{4\gamma} \left \langle  \left \{ \partial_x \Big[ U_1(x)-U_0(x) \Big] \right \} ^2 \right \rangle_{s}, \label{eq:hidden}
\end{equation}
where  $\langle \cdot \rangle_{s}$ denotes the average over the reduced distribution $P^{s}(x)$.     
In the case of a harmonic potential, Eq. (\ref{eq:hidden}) yields
$\dot{W}=k^2 L^2/(2 \epsilon \gamma + 4 \gamma )$. As shown in Fig. \ref{fig:violation-spectrum}(b),
the energy injection rate is always positive and approaches a constant in the limit $\epsilon\to 0$.

We now examine dissipation of the injected energy through viscous relaxation of the bead position $x$, which
is our slow variable. To use Eq.~(\ref{eq:HS}) to compute the associated entropy production,
we need to work out the velocity correlation spectrum $\tilde{C}_{\dot{x}}(\omega)$ 
and the response spectrum $\tilde{R}_{\dot{x}}'(\omega)$. 
It turns out that, in the harmonic case, Eq.~(\ref{eq:noise}) takes a linear form and can be solved
analytically. Here, $\partial_x U_{\sigma_t}(x)=\partial_x U_e(x)-\xi(t)$ is decomposed into an effective force
$\partial_x U_e(x)=k(x-L/2)$ and a ``switching noise'' $\xi(t)=kL(\sigma_t-1/2)$,
with $\langle\xi(t)\rangle=0$ and $\langle\xi(t)\xi(t')\rangle=(kL/2)^2\exp(-2r|t-t'|)$.     
As shown in Fig. 1(c), 
$\tilde{R}'_{\dot{x}}(\omega)=\gamma\omega^2/(k^2+\gamma^2\omega^2)$ is independent of $\epsilon$.  
The correlation spectrum $\tilde{C}_{\dot{x}} (\omega)$, on the other hand,
contains a term $2T\tilde{R}_{\dot{x}}'(\omega)$ from the thermal noise $\eta(t)$,  
and the remaining part from the switching noise $\xi(t)$.
The latter is precisely the FRR violation spectrum shown in Fig. 1(d):
\begin{eqnarray}
 \tilde{C}_{\dot{x}}(\omega)-2T\tilde{R}_{\dot{x}}'(\omega)= \epsilon {k(L/2)^2\over 1+(\omega\tau_f/2)^2} \tilde{R}_{\dot{x}}'(\omega).
 \label{eq:violation}
 \end{eqnarray} 
Carrying out the integral over $\omega$ in Eq.~(\ref{eq:HS}), we obtain $J_x=\dot{W}$. 
Therefore, the FRR violation spectrum of the slow variable $\dot{x}$ allows full recovery of entropy production
in the present case.

A remarkable feature of the FRR violation spectrum displayed in Fig.~1(d), which is also evident from Eq.~(\ref{eq:violation}), 
is the plateau behavior in the intermediate frequency
range $\tau_s^{-1}  \ll  \omega  \ll \tau_f^{-1}$.   
This is the time or frequency window over which the fluctuating dynamics of $x$ deviates most from equilibrium,
and also where the hidden entropy production takes place in the model. 
As $\epsilon$ approaches zero, the height of the plateau diminishes, leaving the apparent impression that
the FRR is restored. Nevertheless, the integral in Eq.~(\ref{eq:HS})
remains finite so as to be consistent with the energy input shown in Fig. 1(b).  
Our explicit solution of the potential switching model thus exposes subtleties surrounding 
the limit $\epsilon\rightarrow 0$.

The above example demonstrates that, with the help of the HS equality, at least part of the hidden entropy production can be
recovered through precise measurement of the FRR violation spectrum of a slow variable.
To gain an impression on the feasibility of this proposal, we have 
explicitly reconstructed the velocity correlation and response spectrum from a simulated stochastic trajectory $x(t)$ of the potential switching model at $\epsilon=0.01$.  The length of the trajectory is taken to be $T_{sp}=10^4\tau_s$, with a sampling rate
of $2\tau_f^{-1}$.  This is sufficient to reveal the full range of the plateau as indicated by open circles in 
Fig.~\ref{fig:violation-spectrum}(d).  
As we show in Supplemental Material~\cite{supp-hiddenEntropy}, the relative fluctuation of $\tilde{C}(\omega)$ goes generally as $(T_{sp}/\tau_s)^{-1/2}$.   
Therefore,  to reach a precision of order $\epsilon$,
the length of the time series should be of the order of $\tau_s/\epsilon^2$.  
Lowering temporal resolution in the measurement will lose information on 
the high frequency end of the spectrum. Nevertheless, even a sampling rate of $0.1\tau_f^{-1}$
can provide good evidence of nonequilibrium fluctuations in $x$ generated by the hidden fast processes.

Below we show that the plateau behavior is a general feature of nonequilibrium Markov systems with timescale separation.
Since a Langevin model can be considered as a special case of the Markov 
process, the aforementioned results can be extended to general potential
switching models with  a nonlinear force field and position-dependent
switching rates, including the well-studied F$_1$-ATPase. 
This constitutes the main result. 

\begin{figure}
\centering
\includegraphics[width=7cm]{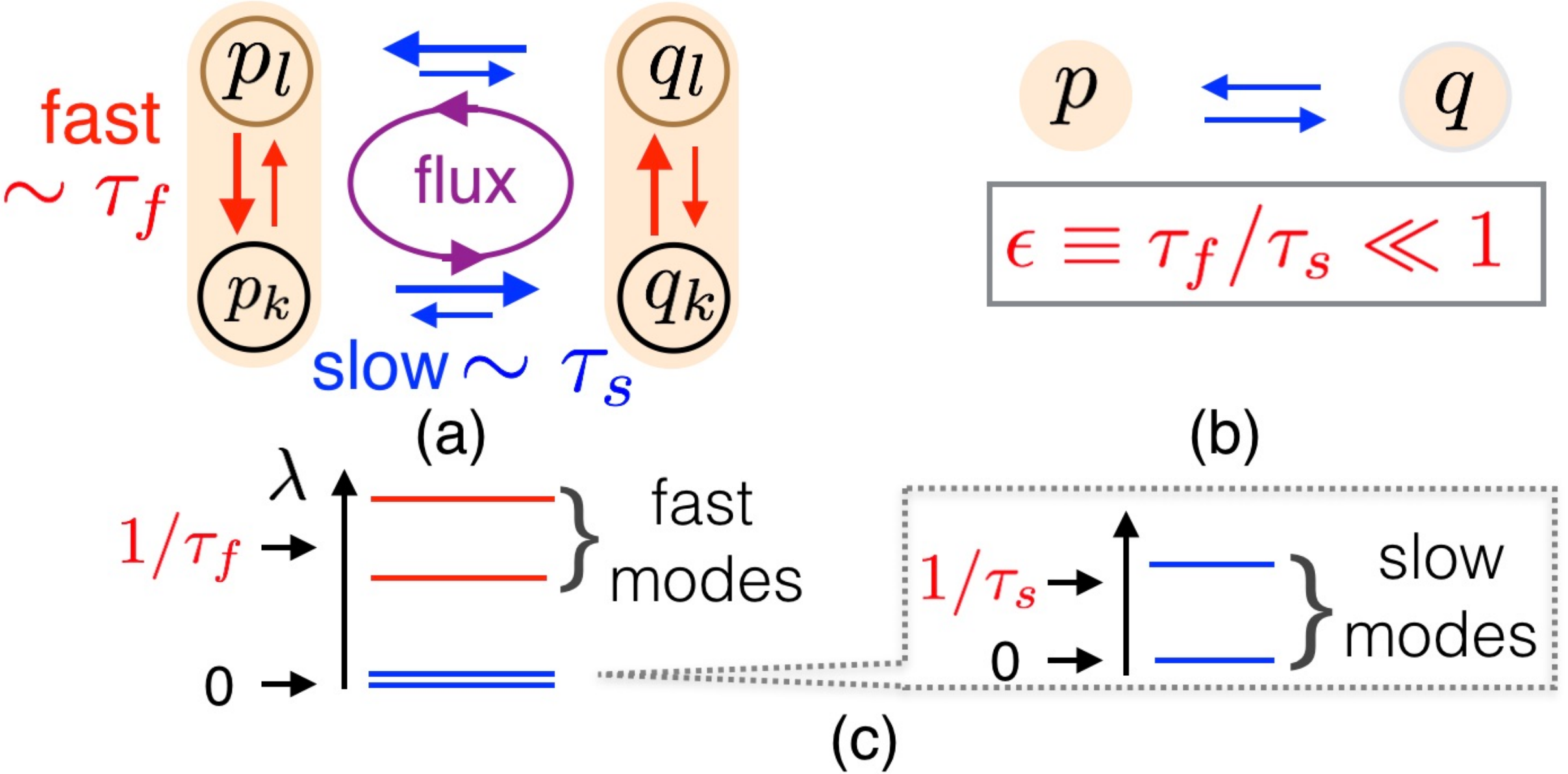}
\caption{(a) Illustration of a general nonequilibrium Markov process with a dissipative cycle formed by fast and slow processes.  
(b) Effective system at large timescale separation $\epsilon\ll 1$,  where the circulating probability flux is hidden, along with the
associated entropy production.  (c) Eigenvalue spectrum of the Master equation for a timescale separated system.
Fast modes are well-separated in their decay rates $\lambda_j$ from the slow ones. The latter form a nearly degenerate
band at the bottom that defines the slow dynamics of the effective system. }
\label{fig:modelSystem}
\end{figure}

\emph{General Markov processes.---}  Consider a general connected Markov system with $N$ states.   
Transition from state $m$ to state $n$ ($1 \leq n,m \leq N$) takes place at rate $w_{m}^{n}$.   
The probability $P_n(t)$ to be at state $n$ after time $t$ follows the Master equation 
\begin{equation}
\frac{d}{dt}P_n(t)=\sum_m M_{nm}P_m(t),
\end{equation}
where  
$M_{nm}=w_m^n-\delta_{nm} \sum_{k}w_n^k $ and $\delta_{nm}$ is the Kronecker delta. %
The right and left eigenmodes, denoted as  $x_j$ and  $y_j$ respectively, satisfy  
$\sum_m M_{nm}x_j (m)=-\lambda_j x_j (n)$ and  $\sum_n y_j (n)  M_{nm}=-\lambda_j y_j (m)$, 
where the minus sign is introduced so that $\operatorname{Re}(\lambda_j)\ge 0$.   
We  rank $\lambda_j$ in an ascending order by its real part, i.e., $\operatorname{Re}(\lambda_1)\le \operatorname{Re}(\lambda_2)\le \cdots$.  The first eigenvalue $\lambda_1=0$,  with $x_1(n)=P_n^{s}$ being the steady-state distribution
and $y_1(n)=1$.  Generically,
the normalised eigenmodes satisfy the orthogonality relation $\sum_n x_j(n)y_{j'}(n)=\delta_{jj'}$ 
and the completeness relation  $\sum_j x_j(n)y_j(n')=\delta_{nn'}$.  

We now introduce an external 
perturbation of strength $h$ whose effect on the dynamics is specified by the modified 
transition rates $\tilde{w}_m^n=w_m^n\exp[(\mathcal{Q}_n-\mathcal{Q}_m)h/2T]$,
where $\mathcal{Q}_n$ is a state variable conjugate to $h$~\cite{ diezemann2005fluctuation, maes2009response}.   
Along a stochastic trajectory $n_t$,  $Q(t)\equiv  \mathcal{Q}_{n_t}$.  In analogy with the velocity variable
for the Langevin dynamics (\ref{eq:noise}), we consider the time derivative $\dot{Q}(t)$ 
whose correlation and dynamic response are defined as 
$C_{\dot{Q}}(t-\tau)\equiv  \langle [\dot{Q}(t)-\langle \dot{Q}\rangle_{s}][\dot{Q}(\tau)-\langle\dot{Q}\rangle_{s}]\rangle_{s}$ and $R_{\dot{Q}}(t-\tau)\equiv \delta \langle \dot{Q}(t)\rangle/\delta h_\tau$, respectively.   
By following the time evolution of the state probabilities $P_n(t)$ under the eigenmode expansion,
we have obtained general expressions for $C_{\dot{Q}}$ and $R_{\dot{Q}}$ which take the following form in frequency,
\begin{subequations}\label{eq:CR-fre}
 \begin{eqnarray}
  \tilde{C}_{\dot{Q}}(\omega)&=&\sum_{j=2}^N 2\alpha_j\beta_j\lambda_j \Big[1-\frac{1}{1+(\omega/\lambda_j)^2}\Big],  \\
 \label{eq:Cvelo_fre}
   \tilde{R}_{\dot{Q}}(\omega) &=&\sum_{j=2}^N \alpha_j\phi_j\Big[1- \frac{1-i(\omega/\lambda_j) }{1+(\omega/\lambda_j)^2}\Big] , 
   \label{eq:Rvelo_fre}
 \end{eqnarray}
\end{subequations}
where  $i$ is the imaginary unit.  
 
The coefficients in Eqs.~(\ref{eq:CR-fre}) are weighted averages of $\mathcal{Q}$, i.e., 
$\alpha_j\equiv \sum_{n} \mathcal{Q}_{n}x_j(n),\; \beta_j \equiv  \sum_{n}\mathcal{Q}_{n}  y_j(n)P_{n}^{s}$, and $\phi_j \equiv  \sum_{n} B_{n} y_j(n)$,  with  $B_n \equiv  \sum_m[ w_m^nP_m^{s}+w_n^mP_n^{s} ](\mathcal{Q}_n-\mathcal{Q}_m)/2T$.  
They satisfy the general sum rule~\cite{shouwen2016PRE},  
\begin{equation}
\sum_{j=2}^N \alpha_j(\lambda_j\beta_j-T\phi_j)=0.
 \label{eq:sum-rule}
\end{equation}
The detailed balance condition $w_n^mP_n^{eq}=w_m^nP_m^{eq}$ implies $\lambda_j\beta_j^{eq}=T\phi_j^{eq} $
for all $j$, and hence the FRR $\tilde{C}_{\dot{Q}}(\omega)=2T\tilde{R}_{\dot{Q}}'(\omega)$.
More generally, in the limit $\omega\rightarrow\infty$, 
$\tilde{C}_{\dot{Q}}(\infty)=2T\tilde{R}_{\dot{Q}}'(\infty)= const$ by virtue of the sum rule, confirming the behavior
seen in Fig.~1(c) on the high frequency side.

We now apply the general results to a timescale separated system whose states can be partitioned 
into $K$ subgroups or coarse-grained states, denoted as $p$ or $q$.  
Each state $n$ ($m$) is alternatively labeled by $p_k$ ($q_l$),  with $k$ ($l$) for a microscopic state within $p$ ($q$). 
Within a coarse-grained state,  transitions are   fast (timescale $\sim \tau_f$),  whereas transitions  across
coarse-grained states are slow (timescale $\sim \tau_s$), as illustrated in Fig.~\ref{fig:modelSystem}(a).   
Formally, this condition amounts to the statement that the transition rate matrix is split into two parts: 
\begin{equation}
M_{p_kq_l}=\epsilon^{-1}\delta_{pq}M^q_{kl}+M^{(1)}_{p_kq_l},
\label{eq:M-separated}
\end{equation}
where $ \epsilon^{-1}M^{q}$ ($\epsilon \equiv  \tau_f/\tau_s$) is the transition rate matrix within a coarse-grained state $q$ 
while $M^{(1)}$  is  for slow transitions between coarse-grained states~\cite{rahav2007fluctuation}.
Below we shall analyze the eigenvalue and FRR spectra of the
Markov process Eq. (\ref{eq:M-separated}) using perturbation theory.

In the absence of inter-group transitions, the matrix $M$ is block diagonalized. Within each block $q$,
the rate matrix $\epsilon^{-1}M^q$ has a nondegenerate eigenvalue $\lambda_1^q=0$ and
the corresponding stationary distribution $P^{s}(l|q)$. All other eigenvalues of the matrix are positive and scale
as $\epsilon^{-1}$,  which together define the fast modes of the system.

Figure~2(c) illustrates the eigenvalue spectrum of the block diagonalized matrix and its modification by $M^{(1)}$
when inter-block transitions are introduced. Lifting of the $K$-fold degeneracy at $\lambda=0$ can
be analyzed using standard perturbation theory. To the leading order in $\epsilon$,
the projected transition rate matrix, 
\begin{equation}
\widehat{M}_{pq}= \sum_{k,l} M_{p_kq_l}^{(1)}P^{s}(l|q),
\end{equation}
defines an emergent dynamics for the coarse-grained states. 
Its eigenmodes $\widehat{x}_j$ and $\widehat{y}_j$ with eigenvalue $\widehat{\lambda}_j$ ($\sim 1$) account for the
leading order behavior of the slow modes $(j\le K)$ in the full state space~\cite{shouwen2016PRE}, which satisfy 
$ x_j(p_k)=\widehat{x}_j(p)P^{s}(k|p)+O(\epsilon)$,
and $ y_j(p_k)=\widehat{y}_j(p)+O(\epsilon)$, and
$\lambda_j=\widehat{\lambda}_j+O(\epsilon)$.  We now focus on a slow observable $Q_{p_k}=Q_p$ that only depends on the coarse-grained state.   Then,  $\widehat{\alpha}_j$, $\widehat{\beta}_j$, and
$\widehat{\phi}_j$ are also defined for the slow modes
$(j \le K)$, which turns out to be
$\alpha_j=\widehat{\alpha}_j+O(\epsilon)$,
$\beta_j=\widehat{\beta}_j+O(\epsilon)$,
and  $\phi_j=\widehat{\phi}_j+O(\epsilon)$.
In terms of the parameters from the coarse-grained dynamics,
we may write
\begin{equation}
 \tilde{C}_{\dot{Q}}-2T\tilde{R}'_{\dot{Q}}=2\sum_{j=2}^K \widehat{\alpha}_j \frac{T\widehat{\phi}_j-\widehat{\beta}_j\widehat{\lambda}_j}{1+(\omega/\widehat{\lambda}_j)^2}+ \epsilon V_s(\omega).
 \label{eq:velo-FRR-vio}
\end{equation}
An explicit form for $V_s(\omega)$ is given in  Supplemental Material~\cite{supp-hiddenEntropy}.
 In the intermediate region ($\tau_s^{-1}\ll \omega\ll\tau_f^{-1}\sim \epsilon^{-1}$),   $V_s(\omega) \simeq 2\epsilon^{-1}\sum_{j=2}^K \alpha_j(\beta_j\lambda_j-T\phi_j)\xrightarrow{\epsilon \to 0} const$ due to the   
sum rule $\sum_{j=2}^K \widehat{\alpha}_j(\widehat{\beta}_j\widehat{\lambda}_j-T\widehat{\phi}_j)=0$ under $\widehat{M}$.  Besides,  $V_s(\omega)$ vanishes for  both $\omega\gg \tau_f^{-1}$ and $\omega\ll \tau_s^{-1}$. 

Equation (\ref{eq:velo-FRR-vio}) is the generalized description of the plateau behavior we found  in the potential switching model
[see Eq.~(\ref{eq:violation})].
The low frequency FRR violation spectrum $(\omega\sim \tau_s^{-1})$ comes from the coarse-grained dynamics, 
which vanishes if the detailed balance condition is fulfilled under $\widehat{M}$.
In the intermediate frequency region,  the non-equilibrium coupling between the fast and slow variables produces a plateau $\epsilon V_s$ whose height diminishes
in the timescale separation limit $\epsilon\rightarrow 0$. 

Since the HS equality holds generally for system variables that follow the Langevin dynamics, Eq.~(\ref{eq:velo-FRR-vio})
can be immediately used to obtain heat dissipation associated with the frictional motion of these variables. On the other hand,
if the slow variable $p$ takes on discrete set of values, then the situation is more complicated.
Lippiello et al.~\cite{Lippiello2014fluctuation} considered a special class of discrete models whose dynamics follows closely
that of a Langevin system.  There it was shown  for a one-dimensional system  that, when the allowed transition rates take the symmetric form
$w_m^n=\tau^{-1}e^{[S(n)-S(m)]/2}$ with $|S(n)-S(m)|\ll 1$, the HS equality holds approximately,  and hence
a link between the FRR violation spectrum and the heat dissipation can again be established.  
In Supplemental Material~\cite{supp-hiddenEntropy}, we present an  example with a ladder network structure.
More general discussion of the relation between the plateau behavior and the hidden entropy production 
is left to future work.

\emph{Concluding Remarks.---}  We have demonstrated in a fairly general setting that, 
for driven systems with large timescale separation, the fluctuation-response relation applied to  
a slow variable is close to be satisfied below its relaxation time. 
However, close examination should reveal
a characteristic plateau behavior in the FRR violation spectrum in the intermediate frequency region. 
 The Harada-Sasa equality can then be invoked to compute the frictional dissipation 
arising from nonequilibrium fluctuations using data from precise measurements of the fluctuation-response spectrum. 

We believe that our findings can be applied to expose hidden entropy production in molecular motors. 
Examples include the F$_{1}$-ATPase
mutants~\cite{toyabe2011thermodynamic,toyabe2015single}, which are known to be less efficient in
converting chemical to mechanical energy as compared to the wild-type.   The fast variable in this case corresponds to the chemical states associated with
ATP binding and hydrolysis, while the relatively slow variable is the rotational angle.  
Their chemomechanical coupling is usually modeled by a potential switching model where position-dependent ATP binding and hydrolysis  shifts the potential forward,  thus generating a directed rotation~\cite{kawaguchi2014nonequilibrium}.          
 Our work suggests that energy loss in inefficient motors may be caused by fast switching of the chemical states 
that produces nonequilibrium fluctuations in the rotary motion.
If so, at least part of the hidden entropy production can be measured by observing the rotary
motion with a high speed camera, without monitoring the
chemical states.  Suppose that the rotory motion has a relaxation time around $ 0.1$s,  and the timescale of ATP binding/hydrolysis around $5\times 10^{-3}$s, then  a sampling duration $T_{sp}=400$s and a temporal resolution $2.5\times 10^{-3}$s,  
which is attainable in state-of-the-art single molecule experiments~\cite{toyabe2012recovery}, would be enough to see the FRR
violation spectrum.   
The rate of ATP binding and ADP release can be modified by varying the concentration of these molecules, yielding further information on the nature of nonequilibrium fluctuations and associated energy dissipation in the system. 
 
\begin{acknowledgements}

The authors thank  Yohei Nakayama and Takayuki Ariga
for helpful suggestions on the manuscript.   
The work was supported in part by the NSFC under Grant No. U1430237 and by the Research Grants Council of the Hong Kong Special Administrative Region (HKSAR) under Grant No. 12301514.  It was also supported by KAKENHI (Nos. 25103002 and 26610115), and by the JSPS Core-to-Core program ``Non-equilibrium dynamics of soft-matter and information''.

\end{acknowledgements}

\onecolumngrid

\def\theequation{S\arabic{equation}}
\makeatletter
\@addtoreset{equation}{section}
\makeatother

\setcounter{equation}{0}

\onecolumngrid
\def\theequation{S\arabic{equation}}
\makeatletter
\@addtoreset{equation}{section}
\makeatother
\setcounter{equation}{0}

\def\thefigure{S\arabic{figure}}
\makeatletter
\@addtoreset{figure}{section}
\makeatother
\setcounter{figure}{0}
\newcommand{\wh}[1]{\widehat{#1}}
\newpage

\section{Supplemental Material}

\subsection{Computation of the correlation and response spectra from data}

To determine the velocity correlation spectrum,  we  do the following: (a) sample the unperturbed trajectory $x(t)$ at a temporal resolution $\delta t$  for a duration $T_{sp}$ ($\gg \tau_s$);  (b) compute the Discrete Fourier Transform (DFT) $\tilde{x}(\omega)$ of $x(t)$, from which the velocity correlation spectrum $\omega^2|\tilde{x}(\omega)|^2 $ is constructed;  (c) following the proposal in~\cite{berg2004power}, perform local average of the spectrum to suppress noise (known as data compression).

To determine the velocity response spectrum at selected frequencies $\{\omega_k\}$ simultaneously,  
we do the following: (a) sample the trajectory $x(t)$ perturbed by a linear combination of periodic driving forces $h=\sum_k h_k\exp(i\omega_k t)$, at a temporal resolution $\delta t$  for a duration $T_{sp}$ ($\gg \tau_s$); (b) compute the DFT $\tilde{x}(\omega)$ of the perturbed trajectory  $x(t)$; (c) at each $\omega_k$, calculate the response 
$\tilde{R}_x(\omega_k)=\tilde{x}(\omega_k)/h_k$,  and thus the velocity response $\tilde{R}_{\dot{x}}(\omega_k)=-i\omega_k \tilde{R}_x(\omega_k)$.
  
Below we examine in some detail fluctuations of the spectra constructed through the above procedure when 
a single long trajectory is used in computation. This will allow us to devise suitable averaging procedures to suppress
noise in the data without significantly hampering its information content.

\subsubsection{Correlation spectrum}

The discussion below follows main ideas of the power spectrum analysis presented in Ref. \cite{berg2004power}.
\vspace{1em}

\noindent \underline{General considerations}
\vspace{1em}

For a general stochastic trajectory $x(t)$ sampled at resolution $\delta t$ and for duration $T_{sp}=N_{sp}\delta t$,  we introduce the following DFT,
\begin{equation}
\tilde{x}(\omega_k)=\frac{\sqrt{\delta t}}{\sqrt{N_{sp}}}\sum_{j=1}^{N_{sp}} x(j\delta t)\exp(-i\omega_k j \delta t) ,
\label{eq:renormalization}
\end{equation}
where $\omega_k=2\pi k/T_{sp}, k=0,\ldots, N_{sp}-1$.  As illustrated by the harmonic potential switching model below, 
each Fourier amplitude $\tilde{x}(\omega_k)$ is a random variable whose value, just like the trajectory $x(t)$, 
changes from realization to realization. Therefore the velocity spectral function
\begin{equation}
g(\omega)=\omega^2|\tilde{x}(\omega)|^2,
\end{equation}
obtained from a single trajectory is a strongly fluctuating quantity.  To estimate the ensemble averaged spectrum
\begin{equation}
\tilde{C}_{\dot{x}}(\omega)=\langle g(\omega)\rangle,
\end{equation}
one may invoke the smoothness of the function by
averaging $g(\omega_k)$ over nearby frequencies in a suitable window of width $\Delta\omega$.
Since the spacing between accessible frequencies is $\delta\omega =2\pi/T_{sp}$, 
relative error of the window-averaged $g(\omega)$ should decrease as 
$(\Delta\omega/\delta\omega)^{-1/2}=(T_{sp}\Delta\omega/2\pi)^{-1/2}$. On the other hand, $\Delta\omega$ should
not be chosen too large to cause significant information loss. For the latter, we may consider the Taylor expansion
of $\tilde{C}_{\dot{x}}(\omega)$ around a chosen $\bar\omega$,
\[\tilde{C}_{\dot{x}}(\omega)=\tilde{C}_{\dot{x}}(\bar{\omega})+{d\tilde{C}_{\dot{x}}\over d\omega}\Bigr|_{\bar\omega}
(\omega-\bar\omega) +\frac{1}{2} \frac{d^2 \tilde{C}_{\dot{x}}}{d\omega^2}\Bigr|_{\bar\omega}
(\omega-\bar\omega)^2 +\ldots\]
Averaging the above expression over a frequency window of width $\Delta\omega$ centered at $\bar\omega$ yields,
\[\overline{\tilde{C}_{\dot{x}}(\omega)}\simeq \tilde{C}_{\dot{x}}(\bar{\omega})+\frac{1}{24} \frac{d^2 \tilde{C}_{\dot{x}}}{d\omega^2}\Bigr|_{\bar\omega}(\Delta\omega)^2 .\]
Equating the two terms on the right-hand-side of the expression above,  we obtain, 
\begin{equation}
\Delta \omega_{\rm M}= \sqrt{24} \left| \frac{\tilde{C}_{\dot{x}}(\bar{\omega})}{\partial^2_{\bar{\omega}}\tilde{C}_{\dot{x}}(\bar{\omega})} \right|^{1/2}.
\end{equation}
This sets an upper bound on the window size for averaging.
In the part of the spectrum where the curvature of $\tilde{C}_{\dot{x}}(\omega)$ is small, 
a large window is desirable so as to reduce statistical error in the original data.

\vspace{1em}

\noindent\underline{Harmonic potential switching model}
\vspace{1em}

We now illustrate the above general ideas with an explicit example, the harmonic potential switching model.  
The bead displacement $x(t)$ from the mid-point $L/2$ satisfies the Langevin equation,
\begin{equation}
\gamma \dot{x}=-kx +\xi(t) +\eta(t),
 \label{eq:Langevin}
\end{equation}
where $\xi(t)$ and $\eta(t)$ are switching and thermal noise, respectively.
Integrating Eq. (\ref{eq:Langevin}) over the time interval $\delta t$, we obtain,
\begin{equation}
x(t+\delta t)=\exp(-\delta t/\tau_s)x(t) +\gamma^{-1}\int_t^{t+\delta t}dt' \exp\bigl({t-t'\over\tau_s}\bigr)[\xi(t') +\eta(t')],
\label{eq:delta_t}
\end{equation}
where $\tau_s=\gamma/k$ is the relaxation time constant of bead displacement.

With the help of Eq. (\ref{eq:delta_t}), and ignoring boundary terms at the beginning and end of the time series,
we obtain the following equation for the DFT $\tilde{x}(\omega)$ when the sampling time $\delta t\ll \tau_s,2\pi/\omega$,
\begin{equation}
i \gamma \omega \tilde{x}(\omega)\simeq -k\tilde{x}(\omega)+\tilde{\xi}(\omega)+\tilde{\eta}(\omega). 
 \label{eq:fourier-eq}
 \end{equation}
Here,  $\tilde{\xi}(\omega)$ and $\tilde{\eta}(\omega)$ are the usual Fourier transforms of 
the active noise $\xi(t)$ and thermal noise $\eta(t)$, respectively, independent of $\delta t$.
Therefore the equation satisfied by $\tilde{x}(\omega)$ is nearly identical to its continuous counterpart,
provided the sampling time interval $\delta t$ is much shorter than both $\tau_s$ and $2\pi/\omega$ (i.e., the part of the
spectrum with $\omega <2\pi/\delta t$).
 
Following Eq. (\ref{eq:fourier-eq}), the velocity correlation spectrum from a single trajectory is given by, 
\begin{equation}
g(\omega)=\omega^2|\tilde{x}(\omega)|^2\simeq{1\over k^2}\frac{\omega^2}{1+ (\omega\tau_s)^2} \left[|\tilde{\xi}(\omega)|^2+ 2\text{Re}\left(\tilde{\xi}(\omega)\tilde{\eta}^*(\omega)\right)+|\tilde{\eta}(\omega)|^2\right],
\label{eq:fluctuating-g}
\end{equation}
where $*$ denotes complex conjugate.   Since the two noises are uncorrelated from each other,
the cross term vanishes upon ensemble average that yields the desired correlation spectrum $\tilde{C}_{\dot{x}}(\omega)$.

The mean of $\tilde{\xi}(\omega)$ and $\tilde{\eta}(\omega)$ are zero while their variances are given by
$\langle|\xi(\omega)|^2\rangle=\epsilon\gamma k(L/2)^2/[1+(\omega\tau_f/2)^2]$ and $\langle|\eta(\omega)|^2\rangle=2\gamma T$,
respectively. On the high frequency end of the spectrum, i.e., $\omega\gg\tau_s^{-1}$, 
$\tilde{C}_{\dot{x}}(\omega)$ varies only on the scale of
$\tau_f^{-1}$ which is much greater than $\tau_s^{-1}$ and even greater than $\delta\omega=2\pi/T_{sp}$. 
Therefore $\Delta\omega$ can
be chosen to be very large to reduce the noise in the bare correlation spectrum (\ref{eq:fluctuating-g}).
On the other hand, for frequencies that are comparable or even smaller than $\tau_s^{-1}$, $\tilde{C}_{\dot{x}}(\omega)$ varies 
appreciably so that the window size is limited by $\tau_s^{-1}$. This is the main reason behind the noisy low frequency
spectrum shown in Fig. 1(d) in the Main Text.

\subsubsection{Fluctuation of the response spectrum}

Now,  we consider applying a periodic driving force $h_0\exp(-i\omega_0 t)$ to the system.  Similarly to Eq.~(\ref{eq:fourier-eq}),  the equation at $\omega=\omega_0$  satisfies,
\begin{equation}
i \gamma \omega_0 \tilde{x}(\omega_0)\simeq -k\tilde{x}(\omega_0)+\tilde{\xi}(\omega_0)+\tilde{\eta}(\omega_0)+\sqrt{T_{sp}} h_0 .
 \end{equation}
 Here, the prefactor $\sqrt{T_{sp}}$ appears due to the rescaling scheme in Eq.~(\ref{eq:renormalization}).    The velocity response spectrum calculated for a single perturbed trajectory is given by,
\begin{equation}
f(\omega_0)\equiv -i\omega_0 \frac{\tilde{x}(\omega_0)}{h_0\sqrt{T_{sp}}}=-\frac{i\omega_0 }{i\gamma \omega_0+k}\left( 1+\frac{ \tilde{\xi}(\omega_0)}{h_0\sqrt{T_{sp}}}+\frac{\tilde{\eta}(\omega_0)}{h_0\sqrt{T_{sp}}}\right),
\label{eq:f-fluc}
\end{equation}
which contains fluctuations.  Its ensemble average converges to the correct result 
\begin{equation}
\langle f(\omega_0)\rangle=  \tilde{R}_{\dot{x}}(\omega_0).
\end{equation}
Its standard deviation  is given by 
\begin{equation}
\sqrt{\langle |f(\omega_0)- \tilde{R}_{\dot{x}}(\omega_0)|^2\rangle}=\frac{1}{h_0\sqrt{T_{sp}}}\tilde{C}_{\dot{x}}(\omega_0).
\label{eq:f-sd}
\end{equation}
Equation (\ref{eq:f-sd}) suggests that fluctuations in the velocity response spectrum can be reduced by increasing $h_0$,  thus the signal/noise ratio, and also by using a longer trajectory.  

However, $h_0$ must be small enough to ensure that we are measuring linear response.  Besides, $T_{sp}$ may also
be limited in experiments.
In such a situation,  one may consider to perform averaging over nearby frequencies to reduce noise in the response spectrum
(\ref{eq:f-fluc}). For example, we may apply a perturbation with multiple frequencies inside a window of width 
$\Delta \omega$, i.e.,   $h_0\sum_{k=-n_r/2}^{n_r/2-1} \exp(i[\omega_0+k \delta \omega] t)$,  where $\delta \omega\gg 2\pi/T_{sp}$ to avoid possible interference effects. Here $n_r=\Delta \omega/\delta \omega$ is the number of frequencies considered. Averaging over the response at these frequencies will further reduce the error (\ref{eq:f-sd}) by a factor $1/\sqrt{n_r}$.

\subsubsection{Error bars on the reconstructed spectra}

\begin{figure}[!h]
\centering
\includegraphics[width=17cm]{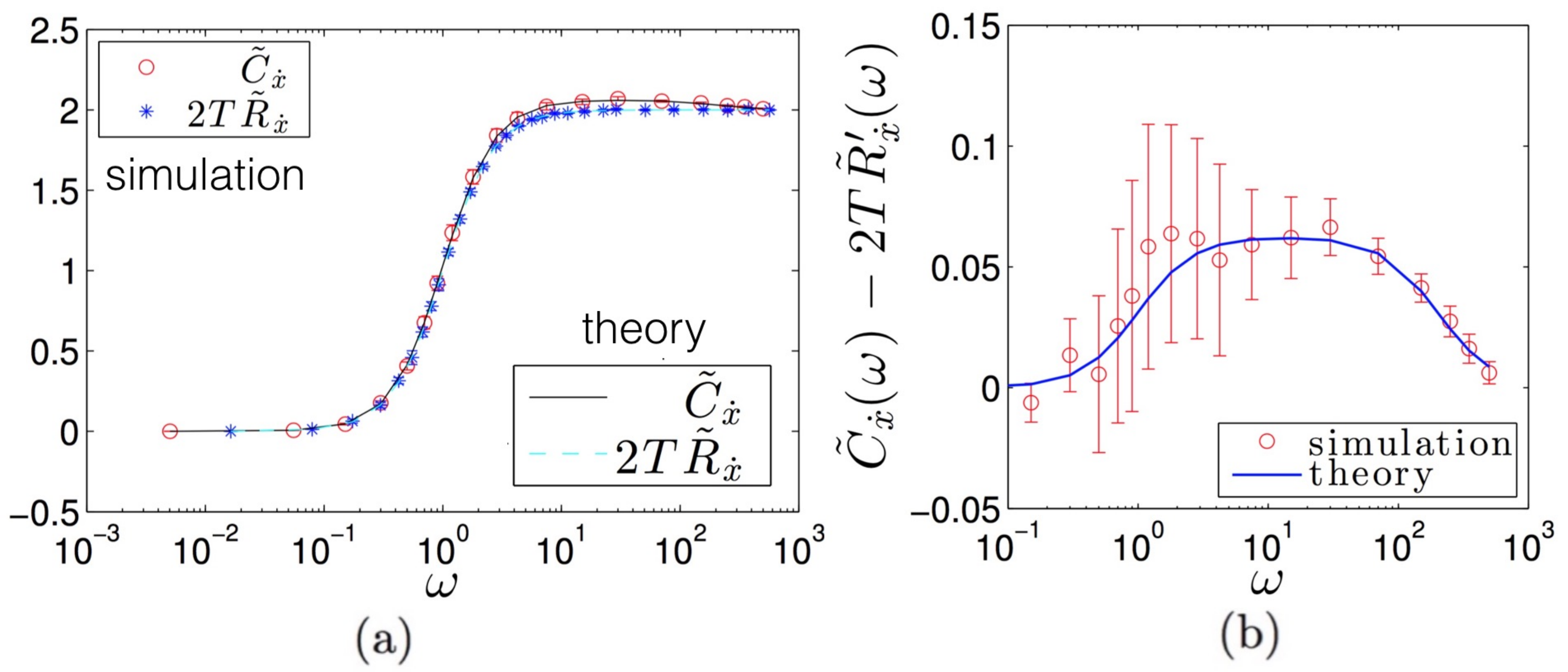}
\caption{ (a) The velocity correlation and response spectrum reconstructed from a simulated trajectory, the same data as those presented in FIG.(1)(c) of the Main Text but with error bars.  (b) The corresponding FRR violation spectrum,  plotted with error bars.  Parameters:  $\gamma=k=1$, $L=5$,   $\tau_f=0.01$,  $T_{sp}=10^4$,   $\delta t=\tau_f/2$, and $h_0=0.5$.  }
\label{fig:error-bar}
\end{figure}

  Fig.~\ref{fig:error-bar} shows error bars on the reconstructed spectra presented in Fig. 1 of the Main Text.   
They have been estimated from variations in the data within each window used for averaging.
For the parameters chosen, our procedure yields very good results on the high frequency end, but less satisfactory
results on the low frequency side close to $\omega\simeq \tau_s^{-1}$. Fluctuations in the latter case are mainly
due to insufficient averaging when computing the correlation spectrum. Although we have not attempted to 
optimize this part of the data analysis, it is conceivable that data smoothening procedures apply to the autocorrelation function
$C_x(t)$ can lead to much improved results.

\begin{figure}[!h]
\centering
\subfigure[]{
\includegraphics[width=8cm]{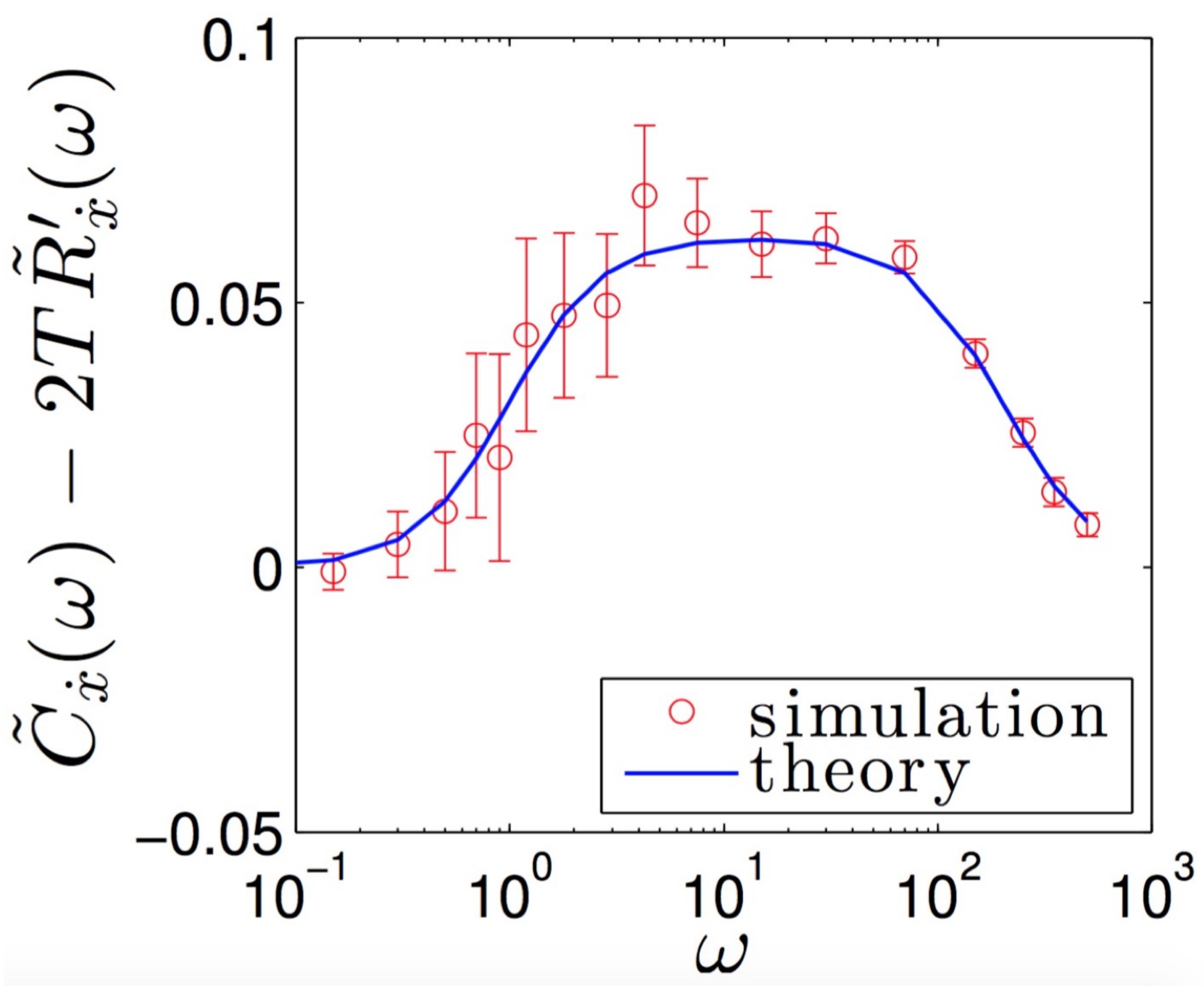}}
\subfigure[]{
\includegraphics[width=8cm]{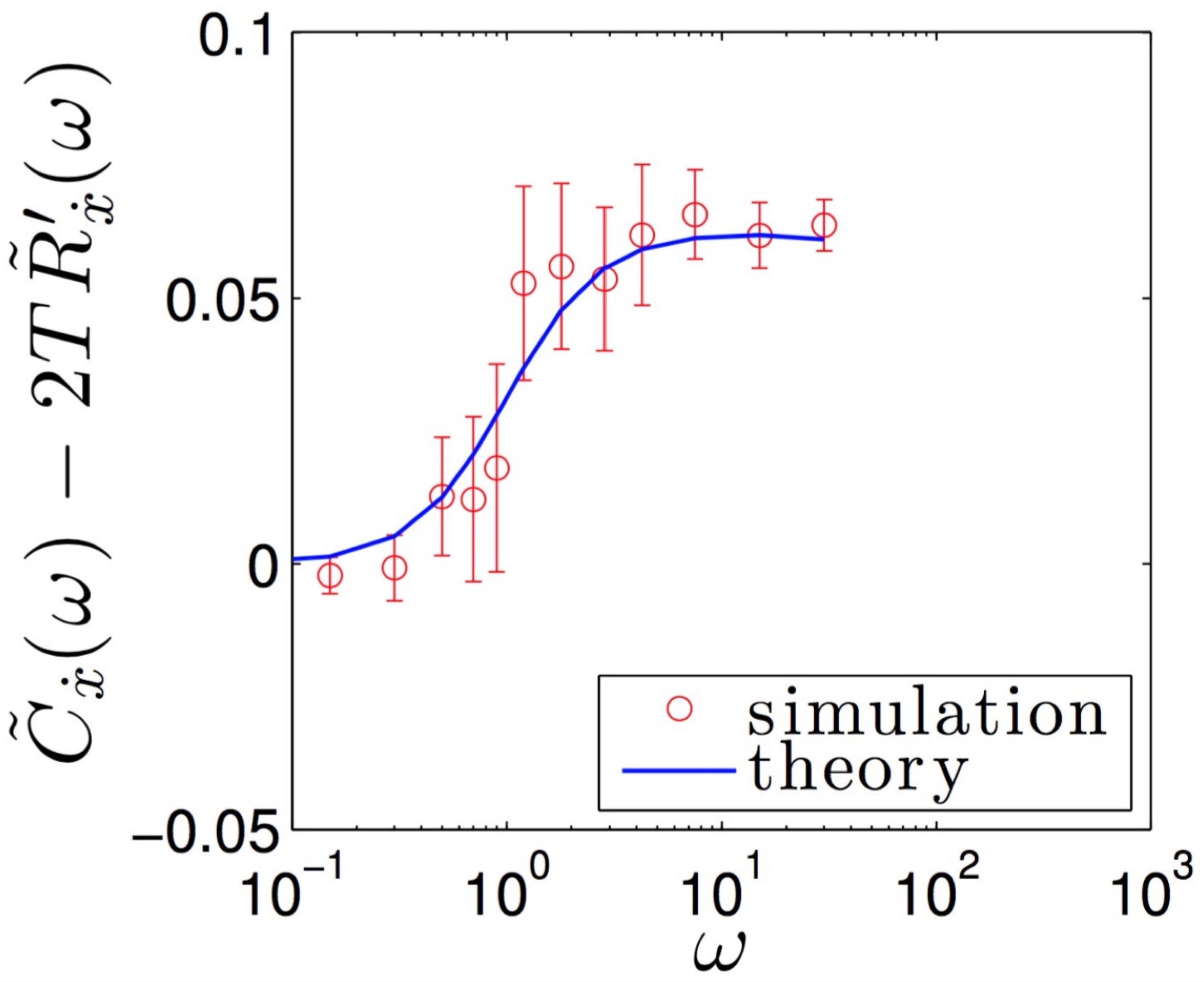}}
\caption{The FRR violation spectrum constructed by using a longer simulation time $T_{sp}=10^5\tau_s$ for the potential switching model. (a) and (b) correspond to the case with temporal resolution $\delta t=\tau_f/2$ and $\delta t=10\tau_f$,  respectively.  Other parameters are the same as FIG.~\ref{fig:error-bar}.  }
\label{fig:L100000}
\end{figure}

Data points in Fig. S2(a) show the numerically determined FRR violation spectrum from a trajectory 10 times 
 longer than the one used to obtain the data points in Fig. S1(b). Fig.~\ref{fig:L100000}(b) shows the data in the same time 
 window but with a much reduced sampling rate
$0.1\tau_f^{-1}$.  The low frequency part of the violation spectrum is faithfully reproduced.

\subsection{Validity of the Harada-Sasa equality in Markov systems}

\begin{figure}[!h]
\centering
\includegraphics[width=16cm]{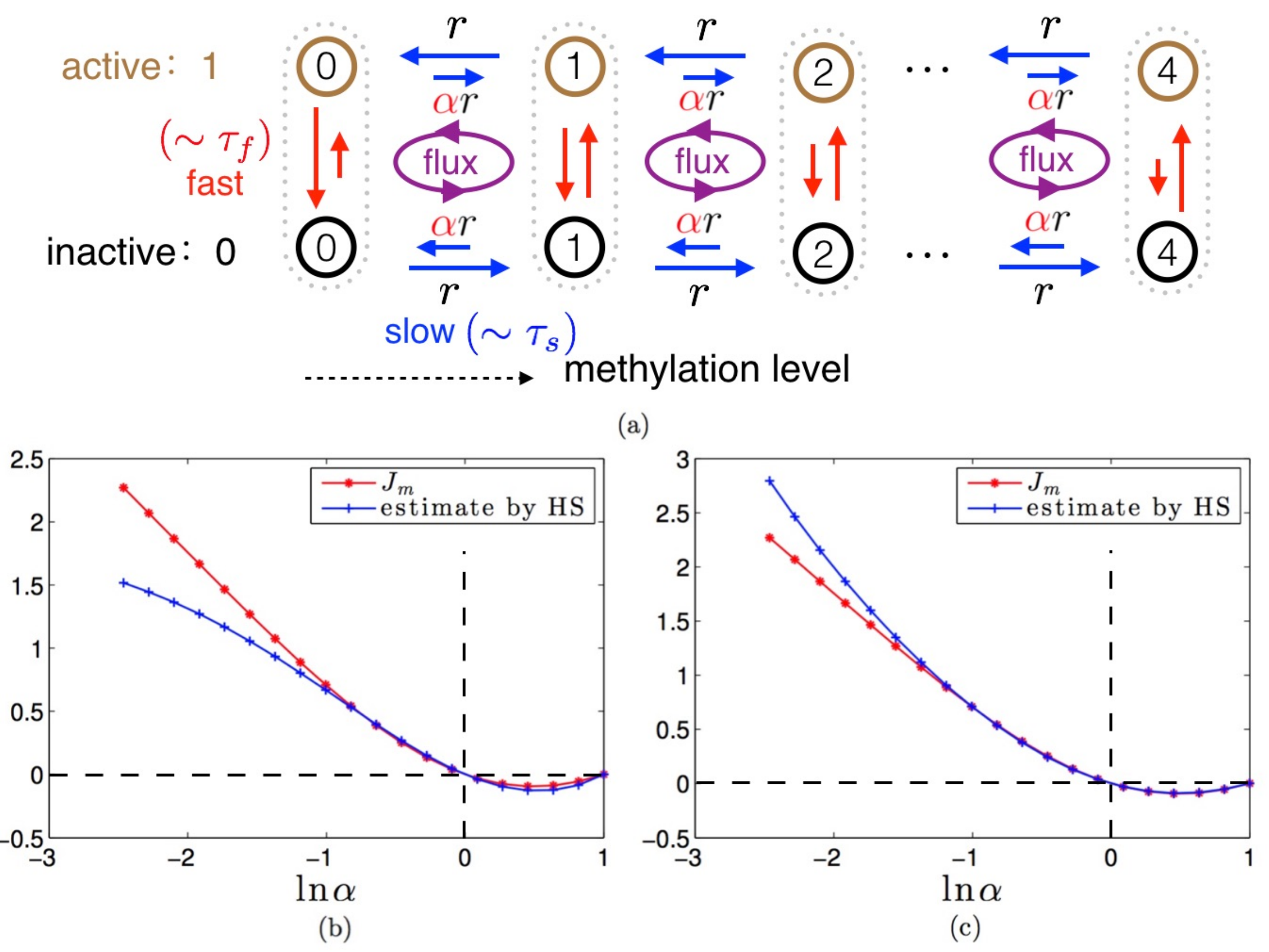}
\caption{(a) The sensory adaptation network for E.coli.  (b)  The comparison between actual dissipation of the methylation dynamics, i.e., $J_m$,  and that estimated from the HS equality,  where we use $\gamma_{eff}=\lim_{\omega\to \infty}1/\tilde{R}_{\dot{m}}(\omega)$. (c) Similar as (b) except that we use   $\gamma_{eff}=T/[r\sqrt{\alpha}]$ here.  }
\label{fig:adaptation}
\end{figure}

In this section,  we show that the HS equality holds approximately for a discrete Markov jump system with a ladder network.  Our argument is a generalization of the work in~\cite{Lippiello2014fluctuation},  where    the HS equality in a one-dimensional discretized Langevin system was studied.  The model we study here,  which  describes the  chemical state dynamics  of the chemotaxis-related membrane chemoreceptor in E.coli~\cite{tu2013quantitative},  is illustrated in FIG.~\ref{fig:adaptation}(a).   The chemical state of this receptor is a combination of the methylation level $m$ ($=0,1,2,3,4$), and the activity $a$ ($=0,1$).    $\alpha$ is a small parameter that tunes the irreversibility of  the methylation/demethylation dynamics, and we only consider the range $\alpha\le \exp(1)$.    The transition from the active state $a=1$ to the inactive state $a=0$ at methylation level $m$ is denoted as $w_1(m)$,  while the reverse transition rate is denoted as   $w_0(m)$.  The rates satisfy
\begin{eqnarray*}
w_0(m)&=&\frac{1}{\tau_f}\exp\left(-\frac{\Delta E(m)}{2T}\right),\\
 w_1(m)&=&\frac{1}{\tau_f}\exp\left(\frac{\Delta E(m)}{2T}\right),
\end{eqnarray*}
where $\Delta E(m)=e_0(m_*-m)$.  Here,  we set $e_0=2$,  $m_*=2$, $\tau_f=0.01$ and $r=1$ in the numerical study.  Therefore,  dynamics of $m$ is relative slow and the timescale separation index $\epsilon\approx 0.01$.   For more details of this model,   refer to ~\cite{shouwen2015adaptation}.  Here,   we study the validity of the HS equality when applied to the slow variable $m$.

Suppose that a transition from state $n$ to $m$ occurs with a rate $w_n^m$.  Then this jump produces entropy $\Delta S_{n}^m=\ln [w_{n}^m/w_{m}^n]$ in the surrounding  media,  and the average dissipation rate of this transition is given by 
\[[P^s_nw_{n}^m-P^s_mw_{m}^n]\ln \frac{w_{n}^m}{w_{m}^n},\]
where $[P^s_nw_{n}^m-P^s_mw_{m}^n]$ is the net flux for this type of transition.   
  The average dissipation rate due to the change of $m$,  denoted as $J_m$,  is obtained by summing over all dissipations that involves changes of $m$,  which   is  given by 
\begin{equation}
J_m= \sum_m\Big( \left[ P^{s}_{1,m+1}r-P^{s}_{1,m}\alpha r \right]\ln \frac{1}{\alpha} +  \left[ P^{s}_{0,m+1}\alpha r-P^{s}_{0,m} r\right]\ln \alpha\Big). 
\label{eq:Jm}
\end{equation}
where $P^s_{a,m}$ is the stationary distribution at state $(a,m)$. 

Now,  we consider how to estimate this dissipation rate from the FRR violation spectrum. 
From the temporal trajectory $m(t)$,  we can compute its velocity correlation spectrum $\tilde{C}_{\dot{m}}(\omega)$ and response spectrum $\tilde{R}_{\dot{m}}(\omega)$.  Then,  we are supposed to estimate the dissipation rate by
\begin{equation}
\gamma_{eff} \left (\langle \dot{m}\rangle+\int_{-\infty}^\infty \left[\tilde{C}_{\dot{m}}(\omega)-2T\tilde{R}_{\dot{m}}'(\omega)\right]\frac{d\omega}{2\pi} \right).
\label{eq:eff-HS}
\end{equation}
The key component in this estimation is the effective friction  coefficient $\gamma_{eff}$, which is in general missing for a Markov process.    Below,  we provide two possible definitions,  both of which are rooted in the generalization of friction  coefficient defined Langevin systems. 

In the over-damped Langevin system,  we know that the friction  coefficient is the inverse of  the velocity response spectrum in the high frequency limit.  Therefore,   we may define   this effective friction  coefficient 
\begin{equation}
\gamma_{eff}=\lim_{\omega\to \infty} 1/\tilde{R}_{\dot{m}}(\omega),
\label{eq:gamma-1}
\end{equation}
  which is directly measurable. 

Alternatively,  if we discretize a continuous over-damped Langevin system with spatial unit $\Delta x$,  its transition rates  take the form 
\begin{equation}
\frac{T}{\gamma \Delta x^2} \exp(\Delta S/2),
\label{eq:DeltaS}
\end{equation}
  where $\Delta S$ is to be understood as the entropy produced in the medium in this single  jump.  The corresponding backward jump produces medium entropy    $-\Delta S$ due to time-reversal asymmetry.   This is consistent with the fact that  more probable transitions  always tend to produce positive entropy in the medium,    which is the microscopic origin for the irreversibility of the second law.    In the case of methylation dynamics in FIG.~\ref{fig:adaptation}, we may  identify $\Delta S$ as the medium entropy produced by the transition with rate $r$,  and  rewrite the methylation/demethylation rates in the form of  Eq.~(\ref{eq:DeltaS}),  i.e., $r=\frac{T}{\gamma_{eff}} \exp(\Delta S/2)$ and $\alpha r=\frac{T}{\gamma_{eff}}  \exp(-\Delta S/2)$,  where the discretization unit is 1 here.   Then,  we can identify  the effective friction  coefficient 
\begin{equation}
\gamma_{eff}=\frac{T}{r\sqrt{\alpha}}.
\label{eq:gamma-2}
\end{equation}

Using the estimator  Eq.~(\ref{eq:eff-HS}) and effective friction  coefficient defined in Eq.~(\ref{eq:gamma-1}),  we obtain the approximate dissipation rate through $m$ dynamics under various $\alpha$,  and compare it with the exact dissipation rate $J_m$,  as shown in FIG.~\ref{fig:adaptation}(b).   It quantitatively captures the  dependence of the dissipation rate on $\alpha$ in the range $\alpha\in [e^{-1} ,\;\;e^1 ]$,  where $|\Delta S|=|\ln \alpha |<1$.    Around $\ln \alpha\approx 0$,  where the two dash lines intersect,  the HS equality becomes almost exact.  Alternatively,   we may use the effective friction  coefficient definition in Eq.~(\ref{eq:gamma-2}),  as shown in FIG.~\ref{fig:adaptation}(c),  which is almost exact in the range  $\alpha\in [e^{-1} ,\;\;e^1 ]$.   Therefore,  both of these two definitions are  correct  in the limit of small entropy production per step, i.e., $|\Delta S|=|\ln \alpha | \ll1$. 

To understand why the HS estimate fails for large $\Delta S$,   below we identify the higher order correction of this estimation.   The HS equality is essentially to  use  the kinetic information  to deduce the energetic information,  or irreversibility of the dynamics.   From the kinetic side, the biased transition rate $(w_+-w_-)$, i.e.,  the difference between forward and backward rate,  provides the hope to extract the dissipation of this single jump, i.e.,  $\ln [w_+/w_-]$.     This is essentially reduced to the following approximation 
\begin{eqnarray}
w_+-w_-&=&\frac{T}{\gamma_{eff}} \exp(\Delta S/2) -\frac{T}{\gamma_{eff}} \exp(-\Delta S/2) \nonumber\\
&=&\frac{T}{\gamma_{eff}} \Big[ \Delta S+ \frac{1}{24}(\Delta S)^3 +o(\Delta S^3)\Big]\nonumber\\
&=&\frac{T}{\gamma_{eff}} \Big[\ln [w_+/w_-]+\frac{1}{24}(\Delta S)^3 +o(\Delta S^3)\Big]
\end{eqnarray}
Therefore,  the HS equality is a leading order approximation,  which becomes exact for Langevin dynamics.  Since  the correction is in the third order of $\Delta S$,  the HS equality can still be reasonably good for many discrete Markov systems,  as we have demonstrated in this sensory adaptation network.   We will discuss this more deeply in our coming paper.

\subsection{Derivation of Equation (10) in the Main Text}

Now,  we study a special type of observable $Q_{p_k}=Q_p$ that only depends on the coarse-grained state, and therefore evolves on a slow timescale $\tau_s$.  We ask how its violation spectrum may reveal the information of the microscopic dynamics that is hidden from our observation.   

Intuitively,  its dynamics seems to be equally well described in both the complete state space by matrix $M$ and the coarse-grained state space by $\wh{M}$.  The projection coefficients for the original system is $\alpha_j\equiv \sum_{p_k} Q_{p}x_j(p_k)$, $\beta_j\equiv \sum_{p_k} Q_{p} y_j(p_k)P^s_{p_k}$,  and $\phi_j\equiv \sum_{p_k} B_{p_k} y_j(p_k)$,  where  $B_{p_k}\equiv \sum_{q_l} [w_{q_l}^{p_k}P^s_{q_l}+w_{p_k}^{q_l}P^s_{p_k}](Q_p-Q_q)/2T$. On the other hand,    in the coarse-grained system,  we may also introduce the projection coefficients $\wh{\alpha}_j\equiv\sum_p Q_p \wh{x}_j(p)$, $\wh{\beta}_j\equiv \sum_p Q_p \wh{y}_j(p) \wh{P}^s_p$,  and $\wh{\phi}_j\equiv \sum_p \wh{B}_p \wh{y}_j(p) $,  where  $\wh{B}_p\equiv \sum_q [\wh{w}_q^p\wh{P}^s_q+\wh{w}_p^q\wh{P}^s_p](Q_p-Q_q)/2T$.  Here,  $\wh{w}_p^q$ is the  transition rate of the coarse-grained dynamics,  which is given by $\wh{w}_p^q\equiv \sum_{k,l}w_{p_k}^{q_l}P^s(k|p)$.  Note that $p\neq q$ here.  Besides,  we define $\lambda_j$ as $\sum_mM_{nm}x_m=-\lambda_jx_n$ and $\sum_my_m M_{mn}=-\lambda_jy_n$,  which affects the sign below. 

To connect the two levels of description, we use the following relations concerning eigenmodes $(j\le K)$ at the two levels,  
\begin{subequations}\label{eq:eigen-xy}
 \begin{eqnarray}
 x_j(p_k)&=&\widehat{x}_j(p)P^{s}(k|p)+O(\epsilon),
 \label{eq:eigen-x}\\
 y_j(p_k)&=&\widehat{y}_j(p)+O(\epsilon),
 \label{eq:eigen-y}
\end{eqnarray} 
\end{subequations}with $\lambda_j=\widehat{\lambda}_j+O(\epsilon)$.  A special but important case of Eq.~(\ref{eq:eigen-xy}) is  
\begin{equation}
P^s_{p_k}=\wh{P}^s_pP^s(k|p)+O(\epsilon).
\end{equation}
Applying these relations to the projection coefficients,  we obtain the following results for the slow modes $(j\le K)$
 \begin{equation}
\alpha_j=\widehat{\alpha}_j+O(\epsilon),\quad \beta_j=\widehat{\beta}_j+O(\epsilon),\quad \phi_j=\widehat{\phi}_j+O(\epsilon).
\label{eq:alpha-relation}
\end{equation}
The sum rule of the effective system $\wh{M}$ demands that 
\begin{equation}
\sum_{j=2}^K \wh{\alpha}_j(\wh{\beta}_j\wh{\lambda}_j-T\wh{\phi}_j)=0.
\label{eq:sum-rule-effective}
\end{equation}

For fast modes $(j>K)$,   perturbative analysis shows that 
\begin{subequations}
  \begin{eqnarray}
  x_j(p_k)&=&\delta_{p,q}\bar{x}_j^q(k)+ O(\epsilon),\\
 y_j(p_k)&=&\delta_{p,q}\bar{y}_j^q(k)+  O(\epsilon),
\end{eqnarray}
  \end{subequations} 
 with $\lambda_j=\epsilon^{-1} \lambda^q+O(1)$.  Here,   $\bar{x}_j^q(k)$ and $\bar{y}_j^q(k)$ are non-stationary eigenmodes of the matrix $\epsilon^{-1}M^q$ describing transitions within the coarse-grained state $q$,  with $\epsilon^{-1}\lambda^q$ being the corresponding eigenvalue.   These non-stationary eigenmodes $(\lambda^q\neq 0)$ satisfy 
 \begin{equation}
 \sum_k \bar{x}_j^q(k)=0,\quad \sum_k \bar{y}_j^q(k)P^s(k|q)=0,
 \end{equation}
 which results from the orthogonal relations with the stationary eigenmodes $(\lambda^q=0)$. 
 With these relations,  we can prove that the fast modes $(j>K)$ satisfy 
   \begin{equation}
\alpha_j=O(\epsilon),\quad \beta_j=O(\epsilon),\quad \phi_j=O(1).
\label{eq:alpha-relation-2}
\end{equation}

With the above preparation,  we now derive Eq.(10) in the Main Text.  First,  we note that there is a big gap between eigenvalues of the slow modes,  which are of order $\tau_s^{-1}\sim 1$,  and those of the fast modes,  which are of order $\tau_f^{-1}\sim \epsilon^{-1}$.  Therefore,  in the frequency region $\omega\ll \tau_f^{-1}$,  only the slow modes contribute to correlation and response spectrum according to Eq.(6) in the Main Text, i.e., 
\begin{subequations}\label{eq:CR-fre}
 \begin{eqnarray}
  \tilde{C}_{\dot{Q}}(\omega)&=&\sum_{j=2}^K 2\alpha_j\beta_j\lambda_j \Big[1-\frac{1}{1+(\omega/\lambda_j)^2}\Big]+\epsilon^3 O([\omega \tau_s]^2),  \\
 \label{eq:Cvelo_fre}
   \tilde{R}_{\dot{Q}}(\omega) &=&\sum_{j=2}^K \alpha_j\phi_j\Big[1- \frac{1-i(\omega/\lambda_j) }{1+(\omega/\lambda_j)^2}\Big]+\epsilon^3 O([\omega \tau_s]^2),
   \label{eq:Rvelo_fre}
 \end{eqnarray}
\end{subequations}
where the correction comes from the fast modes.  According to Eq.~(\ref{eq:alpha-relation-2}),  this correction is of order $ \epsilon (\omega \tau_f)^2$,  or equivalently, $\epsilon^3 (\omega\tau_s)^2$. 
  Then,   the violation spectrum for $\omega\ll \tau_f^{-1}$ can be written as 
\begin{equation}
\tilde{C}_{\dot{Q}}(\omega)-2T\tilde{R}_{\dot{Q}}'(\omega)=2\sum_{j=2}^K \alpha_j\frac{T\phi_j-\beta_j\lambda_j}{1+(\omega/\lambda_j)^2}+2\sum_{j=2}^K \alpha_j(\beta_j\lambda_j-T\phi_j)+\epsilon^3 O([\omega \tau_s]^2).
\label{eq:violation-1}
\end{equation}
Second,  we apply the relations Eq.~(\ref{eq:alpha-relation}) to this violation spectrum,  and obtain  for $\omega\ll \tau_f^{-1}$
\begin{equation}
\tilde{C}_{\dot{Q}}(\omega)-2T\tilde{R}_{\dot{Q}}'(\omega)=2\sum_{j=2}^K \wh{\alpha}_j\frac{T\wh{\phi}_j-\wh{\beta}_j\wh{\lambda}_j}{1+(\omega/\wh{\lambda}_j)^2}+\epsilon V_s(\omega),
\label{eq:violation-2}
\end{equation}
which is exactly Eq.(10) in the Main Text,  with  $\epsilon V_s(\omega)$  a residual contribution defined by Eq.~(\ref{eq:violation-2}).  $V_s(\omega)$ vanishes for $\omega\gg \tau_f^{-1}$ due to FRR in the high frequency limit.  For $\omega\ll \tau_f^{-1}$,  we have 
\begin{equation}
V_s(\omega)=\frac{2}{\epsilon} \sum_{j=2}^K \alpha_j(\beta_j\lambda_j-T\phi_j)+\frac{2}{\epsilon}\sum_{j=2}^K  \alpha_j\frac{T\phi_j-\beta_j\lambda_j}{1+(\omega/\lambda_j)^2}-\frac{2}{\epsilon}\sum_{j=2}^K \wh{\alpha}_j\frac{T\wh{\phi}_j-\wh{\beta}_j\wh{\lambda}_j}{1+(\omega/\wh{\lambda}_j)^2}+\epsilon^2 O([\omega \tau_s]^2). 
\label{eq:Vs}
\end{equation}
The diverging contribution from the first term vanishes due to the approximate  relations Eq.~(\ref{eq:alpha-relation}) and the sum rule Eq.~(\ref{eq:sum-rule-effective}),  and the other two diverging contributions     from the second and the third term cancel each other due to Eq.~(\ref{eq:alpha-relation}).  Therefore,  $V_s(\omega)$ is a well-defined frequency-dependent function  in the timescale separation limit $\epsilon\to 0$.    Below,  we analyze its frequency dependence. 
 
In the intermediate frequency region $\tau_s^{-1}\ll\omega\ll \tau_f^{-1}$, both the second and  the third term in Eq.~(\ref{eq:Vs}) vanish,  and we obtain 
\begin{equation}
V_s(\omega) = \frac{2}{\epsilon}\sum_{j=2}^K \alpha_j(\beta_j\lambda_j-T\phi_j)\xrightarrow{\epsilon \to 0} const
\end{equation}
due to Eq.~(\ref{eq:alpha-relation}) and the   
sum rule Eq.~(\ref{eq:sum-rule-effective}).   For $\omega\ll \lambda_2\sim \tau_s^{-1}$,  the third term in Eq.~(\ref{eq:Vs}) becomes $2\epsilon^{-1} \sum_{j=2}^K\wh{\alpha}_j(T\wh{\phi}_j-\wh{\beta}_j\wh{\lambda}_j)$,  which vanishes due to the sum rule Eq.~(\ref{eq:sum-rule-effective}).  Besides,  the second term becomes $2\epsilon^{-1} \sum_{j=2}^K\alpha_j(T\phi_j-\beta_j\lambda_j)$ in this frequency region,  which cancels the first term.  Therefore,   $V_s(\omega)=0$ for $\omega\ll \tau_s^{-1}$.  To conclude,  $V_s(\omega)$ vanishes both in the high ($\omega\gg\tau_f^{-1}$) and low ($\omega\ll \tau_s^{-1}$) frequency region,  and it becomes a plateau in the intermediate frequency region.

\end{document}